\documentclass[12pt]{article}
\usepackage{a4}
\usepackage{bbm}
\usepackage{latexsym}
\usepackage{amscd}

\renewcommand{\b}{\langle}
\renewcommand{\k}{\rangle}
\renewcommand{\i}{{\rm i}}

\newcommand{\DS}{\displaystyle}

\renewcommand{\d}{{\rm d}}

\renewcommand{\v}[1]{\mbox{\boldmath $#1$}}

\renewcommand{\c}[1]{{\cal #1}}

\newcommand{\bZ}{\mathbbm{Z}}
\newcommand{\bN}{\mathbbm{N}}
\newcommand{\bC}{\mathbbm{C}}
\newcommand{\bR}{\mathbbm{R}}

\newcommand{\End}{{\rm End\,}}
\newcommand{\conv}{{\rm conv\,}}
\newcommand{\Span}{{\rm span\,}}

\newcommand{\bCc}{\mathbbm{C}^{\times}}

\newcommand{\tG}{\tilde{G}}
\newcommand{\cT}{{\cal T}}

\newcommand{\hW}{\hat{W}}
\newcommand{\pf}{{\it Proof.}\hspace{2ex}}
\newcommand{\qed}{\hspace{2em}$\Box$\vspace{1em}}

\newcommand{\eq}[1]{(\ref{#1})}
\newcommand{\pt}[1]{\ref{#1}.}

\renewcommand{\O}{\c{O}}

\newcommand{\C}{\c{C}}

\newcommand{\D}{\c{D}}

\renewcommand{\S}{\c{S}}

\renewcommand{\H}{\c{H}}

\newcommand{\Hu}{{\rm Hu}}
\newcommand{\GG}{{\rm GG}}

\newcommand{\THuc}{\cT_{\Hu}^{\C}}
\newcommand{\THud}{\cT_{\Hu}^{\D}}
\newcommand{\THudp}{\cT_{\Hu}^{\D'}}

\newcommand{\TGG}{\cT_{\GG}^{\O}}
\newcommand{\TGGd}{\cT_{\GG}^{\D}}
\newcommand{\TGGp}{\cT_{\GG}^{{\O}'}}

\newcommand{\WHu}{W_{\Hu}^{\D}}
\newcommand{\WGG}{W_{\GG}^{\D}}

\newcommand{\WGGd}{W_{\GG}^{\D{\textstyle *}}}

\newcommand{\hWHu}{\hW_{\Hu}^{\D}}
\newcommand{\hWGG}{\hW_{\GG}^{\D}}
\newcommand{\hWHud}{\hW_{\Hu}^{\D{\textstyle *}}}
\newcommand{\hWGGd}{\hW_{\GG}^{\D{\textstyle *}}}

\newcommand{\Fc}{F^{\C}_k}

\newcommand{\Fd}{F^{\D}_k}

\newcommand{\Fo}{F^{\O}_l}

\newcommand{\hF}{\tilde{F}}

\newcommand{\hFd}{\hF^{\D}_k}
\newcommand{\hFo}{\hF^{\O}_l}

\newcommand{\gak}{\gamma_k}

\newcommand{\Fcd}{F^{\C{\textstyle *}}_k}

\newcommand{\Fdd}{F^{\D{\textstyle *}}_k}

\newcommand{\hFdd}{\hF^{\D{\textstyle *}}_k}

\newcommand{\Mc}{M^k_{\C}}

\newcommand{\Md}{M^k_{\D}}
\newcommand{\MC}{M^k_{\bC}}

\newcommand{\Mo}{M^l_{\O}}

\newcommand{\gd}{g_l^{\D}}
\newcommand{\gdd}{g^{\D{\textstyle *}}_l}
\newcommand{\go}{g_l^{\O}}

\newcommand{\ec}{e_k^{\C}}
\newcommand{\ed}{e_k^{\D}}

\newcommand{\edi}{e_{k_i}^{\D}}

\newcommand{\Gck}{G^{\C}_k}

\newcommand{\Gdk}{G^{\D}_k}

\newcommand{\Gc}{G^{\C}}

\newcommand{\Gd}{G^{\D}}

\newcommand{\Bp}{B^{\circ}}
\newcommand{\Bpc}{\Bp_{\C}}
\newcommand{\Bpd}{\Bp_{\D}}
\newcommand{\tBpc}{\tilde{B}^{\circ}_{\C}}

\newcommand{\vzd}{\v{z}}
\newcommand{\vz}{\v{\zeta}}

\newcommand{\tfj}{\tilde{f}_j}

\newcommand{\vt}{\tilde{v}}
\newcommand{\wt}{\tilde{w}}

\newcommand{\zt}{\tilde{z}}
\newcommand{\Bt}{\tilde{B}}
\newcommand{\at}{\tilde{a}}
\newcommand{\bt}{\tilde{b}}
\newcommand{\ut}{\tilde{u}}
\newcommand{\mut}{\tilde{\mu}}
\newcommand{\Btp}{\tilde{B}^{\circ}}
\newcommand{\dt}{\tilde{.}}
\newcommand{\Yb}{\overline{Y}}

\renewcommand{\S}{\scriptstyle}

\begin{document}
\thispagestyle{empty}
\hfill
\parbox[t]{3.6cm}{
hep-th/0210215 \\
HD-THEP-02-10 \\
AEI-2002-085 \\
PAR-LPTHE/02-44 \\
PITHA 02/13 \\}
\vspace{2ex}

\begin{center}
\bf\LARGE Topologizations of \\ Chiral Representations
\end{center}

\renewcommand{\thefootnote}{\fnsymbol{footnote}}
\begin{center}
\large Florian Conrady$^{1,2,}\footnotemark[2]$ and Christoph Schweigert$^{3,4,}\footnotemark[3]$
\end{center}
{\small\it
\begin{center}
$^1$Institut f\"{u}r Theoretische Physik, Universit\"{a}t Heidelberg, \\ Philosophenweg 16, D-69120 Heidelberg, Germany \\
$^2$Max-Planck-Institut f\"{u}r Gravitationsphysik, Albert-Einstein-Institut, \\ Am M\"{u}hlenberg 1, D-14476 Golm, Germany
\end{center}
\begin{center}
$^3$LPTHE, Universit\'{e} Paris VI, 4 place Jussieu, \\ F-75252 Paris Cedex 05, France \\
$^4$Institut f\"{u}r Theoretische Physik, RWTH Aachen, \\ Sommerfeldstr. 28, D-52074 Aachen, Germany
\end{center}}
\footnotetext[2]{florian.conrady@roma1.infn.it}
\footnotetext[3]{schweigert@math.uni-hamburg.de}
\renewcommand{\thefootnote}{\arabic{footnote}}
\setcounter{footnote}{0}
\begin{abstract}
\noindent
We analyze and compare two families of topologies that have been proposed
for representation spaces of chiral algebras by Huang and Gaberdiel \&
Goddard respectively. We show, in particular, that for suitable pairs the
topology of Gaberdiel \& Goddard is coarser. We also give a new proof that
the chiral two-point blocks are continuous in the topology of Huang.
\end{abstract}
\setlength{\jot}{2.5mm}
\section{Introduction and Summary}
\label{Introduction_and_Summary}
Two-dimensional
conformal field theories (CFTs) play a key role in the worldsheet
formulation of string theory and in the description of universality classes of critical phenomena.
In the attempt to gain a better
understanding of their mathematical structure, several axiomatic approaches
have been developed.
When using an operator calculus,
a space of states $V$ has to be specified that consists of
representations
of the symmetry algebra. The representation spaces form the basic structure of $V$, but
it is not fully determined by physical requirements
what topology should be given to this space, and hence how it should be completed.
In unitary CFTs, one has, by definition, a positive definite inner product
$\b\;,\;\k$; the conventional approach
is then to use the norm
$\|.\| = \sqrt{\b\;,\;\k}$ for completing $V$ so that one obtains a Hilbert
space $\H$.
Problems usually arise from the fact that domains and ranges for different
operators do not coincide.
Special care has to be taken when considering
operator products. Often this aspect is left aside and
one works on
the {\it assumption} that domain issues can be settled.

To define convergence in terms of an inner product is by no means the only
possibility, nor is it clear that it provides the best starting point for
dealing satisfactorily with domain questions. The theory of distributions
shows that it can prove extremely useful to introduce topologies different
from that of a (pre-) Hilbert space. It was
especially B\"{o}hm who argued for the application of such topologies
to quantum theory (see \cite{Madrid}, sec.\ 1 for references). More recently
this idea has reappeared in the context of CFT when Gaberdiel \& Goddard
\cite{Gaberdiel}
and Huang \cite{Huang} proposed new topologies for
representation spaces of chiral symmetry algebras. A central role is played
by chiral (or conformal) blocks whose properties lead to the definition of
locally convex state spaces that are not Hilbert spaces.
In both cases, one deals with a whole family of topologies which are
parametrized by suitable subsets of the complex plane. In this article we
investigate and compare the two approaches.

Questions of topology
naturally occur when it
comes to constructing a mathematically rigorous operator formalism.
Gaberdiel, Goddard and Huang have achieved that for chiral CFT on
the Riemann sphere. One hopes that these results can be
further generalized, and that the topological properties of the
spaces help in deriving statements that could not be proven so
far. Let us mention a few possibilities:
\begin{itemize}
\item
String theory makes it necessary to deal with CFTs on surfaces of
arbitrary genus $g$. For $g>1$ and interacting theories,
an operator formalism has yet to be developed.
In a different approach one defines conformal blocks as
linear functionals on tensor products of representation spaces \cite{Frenkel}.
The nuclear mapping theorem for nuclear spaces could
provide a way to construct vertex operators from these functionals.
\item
The space of physical superstring states $V_{\rm phys}$ is obtained by
taking the BRST cohomology of the combined matter-ghost system.
The choice of topology can influence
the content of $V_{\rm phys}$, and should
play a role in the construction of picture-changing operators \cite{Berkovits}.
\item
A cohomological approach to the Verlinde formula has been advocated in
the literature \cite{Teleman,Fuchs2}.
The correct dimension of chiral blocks
is obtained provided a certain sequence of coinvariants is exact.
This may be easier to show if one chooses a suitable, possibly nuclear,
topology on the state space.
\end{itemize}
The results of this paper contribute to a
better understanding of the topologies given in
\cite{Gaberdiel} and \cite{Huang}. As they were defined
in rather different axiomatic settings, we have translated
them into a
common framework that allows for the construction of both types of topologies. This
framework is specified in the next section. Section
\ref{Gaberdiels_and_Goddards_Topology} and \ref{Huangs_Topology} explain the definition
of the topologies: they are denoted by $\TGG$ and $\THud$ and parametrized by open sets
$\O\subset\bC$ and open disks $\D\subset\bC$ centered at 0 respectively.
In section \ref{Properties_of_the_Topologies}, we derive some
simple properties of $\TGG$ and $\THud$:
it is shown that $\THud$ is nuclear and that $\TGG$
is nuclear if it is Hausdorff. We also take a look
at the $\O$- and $\D$-dependence:
$$\mbox{$\TGGp$ is coarser than $\TGG$, if $\O'\subset\O$,}$$
whereas $\THud$ behaves in the opposite way:
$$\mbox{$\THudp$ is finer than $\THud$ for $\D'\subset\D$.}$$
Section \ref{Comparison_of_the_Topologies} deals with
the comparison of the topologies. We prove that
$$\mbox{$\THud$ is finer than $\TGG$\hspace{2mm}if\hspace{2mm}$\DS\inf_{\zeta\in\O} |\zeta| > r$,}$$
where $r$ is the radius of the disk $\D$. The techniques employed
allow us to show in section \ref{Continuity_of_Conformal_Blocks}
that conformal two-point blocks of the vacuum sector are
continuous in $\THud$ if the radius of $\D$ is less than half the
distance of the two points. This result generalizes to an
arbitrary number of points.
(One can give another proof of continuity which is based on theorem 2.5 in \cite{Huang}.)
Section 6 and 7 can be read
independently of section 5. In the last section, we make some comments on open questions and speculate on the possibility that Huang's and Gaberdiel's \& Goddard's topologies are dual to each other.

We assume that the reader is familiar with the basics
of the theory of vertex algebras
and the theory of topological vector spaces.
The necessary background material can be found in
\cite{Kac,Frenkel} and
\cite{Narici,Treves}.
\subsubsection*{Acknowledgements}
This work is part of F.\ Conrady's diploma thesis and was
supported by the ``Schwer\-punktprogramm String-Theorie'' of the
Deutsche Forschungsgemeinschaft. We are grateful to M.G.\ Schmidt
and J.\ Fuchs for their continuous interest in this work. We thank
Y.-Zh.\ Huang for helpful correspondence.
\section{Mathematical Framework}
\label{Mathematical_Framework}
The state space of
a CFT is built from representation spaces of a chiral symmetry algebra. The
work of Huang and Gaberdiel \& Goddard provides
us with methods to topologize such spaces.
This section fixes the definition of the chiral representations and specifies the additional
assumptions needed for their topologization.
\subsection{Vertex Algebras and Vertex Algebra Modules}
In this article, we use the concept of vertex algebras
to formally define the chiral symmetry algebra
\cite{FLM,Kac,Frenkel}. The topologies
will be defined on certain vertex algebras and
modules of them.

Let $V$ be a $\bZ_+$-graded vertex algebra consisting of finite-dimensional
graded components, that is
$$V = \bigoplus_{h\in\bZ_+} V_h$$
and
$$\dim V_h < \infty\,.$$
The vacuum vector
is denoted by $\Omega$ ($\Omega\in V_0$).
The map
$$
\renewcommand{\arraystretch}{1.5}
\begin{array}{llll}
\phi: & V & \to & \End V\,[[z,z^{-1}]]\,, \\
& v & \mapsto & \DS \phi(v,z) = \sum\limits_{n\in\bZ} (v)_n\, z^{- n - 1}\,,
\end{array}
$$
establishes the state-field correspondence.
In physical jargon, the endomorphisms
$(v)_n$ are called mode operators of the
field $\phi(v,z)$ associated
to the state $v$.
(The use of the letter $\phi$ is conventional in
quantum field theory; $Y$ is the standard symbol used in the theory of
vertex algebras.)
In a conformal vertex algebra, the grading corresponds to the
assignment of conformal weights.
The choice of an integer grading means that we only consider
bosonic fields; the restriction to positive values follows from
unitarity requirements (see below).

Take $W$ to be a $\bR_+$-graded $V$-module \cite{Frenkel}
satisfying
$$W = \bigoplus_{h\in\bR_+} W_h$$
and
$$\dim W_h < \infty\,.$$
The fields of the chiral algebra $V$ are represented on $W$ by
$$
\renewcommand{\arraystretch}{1.5}
\begin{array}{llll}
\phi_W: & V & \to & \End W\,[[z,z^{-1}]]\,, \\
& v & \mapsto & \DS \phi_W(v,z) = \sum\limits_{n\in\bZ} (v)_n\, z^{- n - 1}\,.
\end{array}
$$
In the remainder of the text the index $W$ is omitted:
it will be clear from the context whether one deals with
fields and operators of the vertex algebra or those of its module.

Let $X\subset V$ and $Y\subset W$
be
subspaces which generate $V$ and $W$
respectively: \setlength{\jot}{4mm}
\begin{eqnarray}
\label{XgeneratingV}
V & = & \Span\{(x_1)_{n_1}\cdots (x_k)_{n_k}x_{k+1}\;|\;x_i\in
X,\,n_i\in\bZ_+,\,k\in\bN\}\,, \\
\label{YgeneratingW}
W & = & \Span\{(v_1)_{n_1}\cdots (v_k)_{n_k}y\;|\;v_i\in
V,\,y\in Y,\,n_i\in\bZ_+,\,k\in\bN\}\,.
\end{eqnarray} \setlength{\jot}{2.5mm}
We assume that $X$ contains $\Omega$. Line
\eq{XgeneratingV} implies that every mode operator $(v)_n$ ($v\in V$, $n\in\bZ_+$) is a linear
combination of products $(x_1)_{n_1}\cdots (x_k)_{n_k}$. 
This can be shown by induction and a suitable
integration of the operator product expansion
(see sec.\ 1.6 of \cite{Frenkel}).
Hence, as a consequence of \eq{XgeneratingV} and \eq{YgeneratingW}, $W$
is spanned by vectors of the form
$$(x_1)_{n_1}\cdots (x_k)_{n_k} y\qquad (k\in\bN)$$
where $x_1,\ldots x_k\in X$, $n_1,\ldots, n_k\in\bZ_+$ and $y\in Y$.

\subsection{Unitarity, Correlation Functions and Finiteness}
\label{Unitarity,_Correlation_Functions_and_Finiteness}
Gaberdiel's and Goddard's axioms lead to
state spaces of a chiral symmetry algebra
that contains at least the M\"{o}bius algebra.
An additional condition on the amplitudes implies the existence of
an inner product and that M\"{o}bius transformations are unitary w.r.t.\ it
(see sec.\ 3.5, \cite{Gaberdiel2}).
In this paper, we assume that this condition is fulfilled,
so effectively one deals with chiral
representations that carry a (pseudo-)unitary structure.

We implement these requirements as follows:
$V$ and $W$ are equipped with inner products $\b\;,\;\k$ (antilinear in the
first variable),
and the Lie algebra ${\rm sl}(2,\bC)$ is unitarily represented on them;
the associated operators $L_0$ and $L_{-1}$ can be identified with the grading
and shift operators respectively\footnote{cf.\ the definition of a
M\"{o}bius-conformal vertex algebra (see, e.g.\ \cite{Kac})}.
It follows that the inner products are compatible
with the grading of $V$ and $W$. Note also that
the inner products can be indefinite; for sake of simplicity, we restrict
ourselves to unitary CFTs and
thus assume that $\b\;,\;\k$ is positive definite.
As a result, the grading of $V$ and $W$ has to be real and positive.

Matrix elements
\begin{equation}
\label{matrixel}
\b\wt, \phi(v_1,z_1)\cdots\phi(v_k,z_k)w\k
\end{equation}
of field products are obtained by inserting
the formal sum
$$\phi(v_1,z_1)\cdots\phi(v_k,z_k)$$
between states $w, \wt\in W$
and replacing the formal variables by complex numbers $z_1,\ldots,z_k$.
In other words, we consider $k+2$ point blocks on the sphere with
in- and out-state taken from the module $W$ and $k$ insertions that are
descendants of the vacuum.
It follows from the axioms of the vertex algebra module
that \eq{matrixel}
converges absolutely in the region
$$|z_1|>\ldots>|z_k|> 0\,,$$
and can be analytically extended to a meromorphic function
on the domain
$$M^k = \{(z_1,\ldots,z_k)\in (\bCc)^k\;|\;z_i\ne
z_j\;\mbox{for}\;
i\ne
j\}\,.$$
$M^k$ is the moduli space of $n$ different ordered points
on $\bCc$.

In the theory of vertex algebras, one frequently considers
``matrix elements'' of the form
\begin{equation}
\label{matrixelgrad}
w'(\phi(v_1,z_1)\cdots\phi(v_k,z_k)w)\,,
\end{equation}
where $w'$ is an element of the graded dual
$$W' = \bigoplus_{h\in\bR_+} (W_h)^*\,.$$
Since the graded components of $W$ are finite-dimensional,
every bra-vector $\b\wt\hspace{0.3mm}|$ can be represented
by some dual vector $w'\in W'$, and all theorems
for matrix elements \eq{matrixelgrad} apply as well to \eq{matrixel}.
For later use, we note here that if $W = V$ and $w = \wt = \Omega$,
the amplitude \eq{matrixel} is translation-invariant:
$$\b\Omega,\phi(v_1,z_1)\cdots\phi(v_k,z_k)\Omega\k
= \b\Omega,\phi(v_1,z_1+z)\cdots\phi(v_k,z_k+z)\Omega\k\,,\quad z\in\bC\,.$$

Given an open set $\D\subset\bC$ we define the space
of ``correlation functions''\footnote{Note that these functions are
objects of the {\it chiral} CFT. They are the chiral (or conformal)
blocks from which the
physical correlators of the full CFT are constructed.}
$\hFd$ $(k\in\bN)$ to be the vector space of all functions
$$\b \wt,\phi(v_1,.)\cdots\phi(v_k,.)w\k\,,$$
with $v_1,\ldots v_k\in V$, $w,\wt\in W$ and arguments $(z_1,\ldots,z_k)$
in the domain
$$\Md = \{(z_1,\ldots,z_k)\in \D^k\;|\; \;z_i\ne
z_j\;\mbox{for}\; i\ne j;\; z_i \neq 0\}\,.$$
$\hFd$ is endowed with the topology of compact convergence,
i.e.\ the topology of uniform convergence on compact subsets of
$\Md$. We denote by $\Fd$ the completion of $\hFd$.
The topological dual $\Fdd$ receives the strong topology --- the topology
of uniform convergence on all weakly bounded subsets of $\Fd$.

For the construction of Huang's topology it is necessary to impose
two additional conditions on the vertex algebra $V$ and the
$V$-module $W$. Both should be {\it finitely generated}:
the spaces $X$ and $Y$ are assumed to be finite-dimensional; we write
$d = \dim X$ and $n = \dim Y$. (Gaberdiel and Goddard only require that $X$
has a countable basis. Various other finiteness conditions have been studied
in the literature, see e.g.\ \cite{Nagatomo}.)

\section{Gaberdiel's and Goddard's Topology}
\label{Gaberdiels_and_Goddards_Topology}
A set of meromorphic and M\"{o}bius covariant amplitudes provides
the starting point for Gaberdiel's and Goddard's definition of chiral
CFT. It allows for a direct construction of vertex operators as continuous
maps between topological spaces. In sections 4 and 8 of \cite{Gaberdiel},
it is explained how this leads to the more common description
in terms of chiral algebras and their representations.
We will not discuss this relation and define the topologies
directly using the vertex algebra $V$ and the module $W$.
Below we give the construction for the module $W$;
it applies in particular to $V$, since $V$ is a finitely generated
module over itself; in this case, $X$ plays the role of $Y$.

Let us first sketch the idea:
we seek to define
seminorms on $W$; to this end, we fix vectors
$\wt\in W$ and $v_1,\ldots,v_l\in V$,
and consider, for each vector $w$, the correlator
$$\b\wt, \phi(v_1,.)\cdots\phi(v_k,.)w\k$$
as a function of $k$ arguments in the complex plane.
After choosing a suitable domain $D$ for these functions,
a seminorm is provided by the supremum norm on compact subsets $K$
of $D$. In other words, one uses the topology of compact
convergence on function spaces in order to topologize
the vector space $W$.
To keep the number of seminorms countable, we restrict
the choice of out-states and insertions to $Y$ and $X$
respectively. That is, we only consider
seminorms of the type
$$\|w\| = \sup_{(\zeta_1,\ldots,\zeta_l)\in K} |\b y,\phi(x_1,\zeta_1)\cdots
\phi(x_l,\zeta_l)w\k|\,.$$

The proofs in the following sections require us to put this scheme
into more formal language:
Let $\O$ be an arbitrary open subset of $\bC$ and $l\in\bN$.
(The spaces $\Mo$ and $\Fo$ were defined in  sec.\
\ref{Unitarity,_Correlation_Functions_and_Finiteness}.)
There is a linear map
$$\go: \Yb\otimes X^{\otimes l}\otimes W\to\Fo$$
defined by
$$\go(y\otimes x_1\otimes\cdots\otimes x_l\otimes w) := \b y,\phi(x_1,.)\cdots \phi(x_l,.)w\k$$
for $y\in\Yb$, $x_1,\ldots,x_l\in X$ and $w\in W$. Here, we use
the complex conjugate space $\Yb$
of $Y$ and thereby avoid the antilinearity of the inner product\footnote{$\Yb$ and $Y$ are identical
as sets and additive groups, only the scalar multiplications
$\bar{\cdot}$ and $\cdot$ differ: they are related by complex conjugation, $a\,\bar{\cdot}\,y = \overline{a}\cdot y \equiv \overline{a}\,y$
for $a\in\bC$.}.
For fixed $l\in\bN$ and
$x\in\Yb\otimes X^{\otimes l}$, we obtain a linear map
\begin{equation}
\label{onlyWF}
\go(x\otimes .): W\to\Fo\,.
\end{equation} 
The family of mappings
$\go(x\otimes .)$, $l\in\bN$, $x\in \Yb\otimes X^{\otimes l}$, determines an {\it initial
topology} $\TGG$ on $W$, i.e.\ the weakest topology with respect to which all
$\go(x\otimes .)$ are continuous.
It is locally convex, but not necessarily
Hausdorff.
At this point, Goddard and Gaberdiel divide out the subspace of
vectors that have zero length with regard to all seminorms and obtain a Hausdorff space. In this paper, we will not do so, as we want to compare topologies on $W$ (and not on some quotient space whose content depends on one of the topologies). We refer to $\TGG$ on $W$ as Gaberdiel's and Goddard's topology, but it should be kept in mind that their {\it space of states} arises only after division by the ``null states''.

Given bases
\begin{equation}
\label{basis}
x_1,\ldots,x_d
\end{equation}
for $X$ and
\begin{equation}
\label{basisY}
y_1,\ldots,y_n
\end{equation}
for $Y$,
multi-indices
$$I = (i_0,i_1,\ldots,i_l)\in \{1,\ldots,n\}\times\{1,\ldots,d\}^l$$
can be used to label a basis
$$x_I = y_{i_0}\otimes x_{i_1}\otimes\cdots\otimes x_{i_l}$$
for $\Yb\otimes X^{\otimes l}$.
By linearity, $\go(x\otimes .)$ is continuous for every
$x\in \Yb\otimes X^{\otimes l}$ iff it is continuous for every
$x_I$.
$\TGG$ is therefore the weakest
topology on $W$ for which each $\go(x_I\otimes .)$, $l\in\bN$,
$I\in\{1,\ldots,n\}\times\{1,\ldots,d\}^l$, is continuous.
It is characterized by the family of seminorms
\begin{equation}
\label{semiGod}
\|w\|_{I,K} := \|\go(x_I\otimes .)\|_K = \|\b y_{i_0},\phi(x_{i_1},.)\cdots
\phi(x_{i_l},.)w\k\|_K\,,
\end{equation}
where the multiindex $I$ specifies the basis element $x_I$ and
$\|.\|_K$ is
the supremum norm on compact subsets $K\subset\Mo$.
This family of seminorms is equivalent to a countable set of seminorms,
since we may restrict our choice of $K$ to a sequence $\{K_n\}_{n\in\bN}$ of compacta
which exhaust $\Mo$.
Note that in the definition we are free to replace
the completion $\Fo$ by $\hFo$ itself without
affecting the topology $\TGG$.

\section{Huang's Topology}
\label{Huangs_Topology}
Huang constructs a topology for finitely generated conformal
vertex algebras and for finitely generated modules associated to them \cite{Huang}.
His formalism
does not rely on the existence of an inner product:
the graded dual is employed for defining matrix elements.
We use the inner product instead and adapt
Huang's scheme
accordingly. The differences are pointed out at the end
of this section. For the complete proofs, we refer the reader
to Huang's paper.

Again, we describe the topology for a finitely generated
module $W$; this includes the specific case $W = V$
where $Y$ is given by $X$.
$\TGG$ was obtained by mapping $W$ into spaces whose topology was already known.
Huang takes the reverse approach: he maps a sequence of
topological spaces into a
vector space containing $W$, and equips it with
the {\it strict inductive limit topology} (for a definition, see e.g.\ \cite{Narici}, chap.\ 12).

We take $\D$ to be an open disk of arbitrary radius $r>0$ around 0.
Let $\overline{\rm Hom}(W,\bC)$ be the space of antilinear functionals on $W$.
Again, the conformal blocks are used as a key input;
we specify a map
\begin{equation}
\label{ed}
\ed: X^{\otimes k}\otimes Y\otimes\Fdd\to\overline{\rm Hom}(W,\bC)
\end{equation}
by
\begin{equation}
\label{ek}
\ed(x_1\otimes\cdots\otimes x_k\otimes y\otimes\mu)(\wt) := \mu(\b\wt, \phi(x_1,.)\cdots \phi(x_k,.)y\k)
\end{equation}
for $x_1, \dots, x_k\in X$, $y\in Y$, $\mu\in\Fdd$ and $\wt\in W$.
Here, $$\b\wt,\phi(x_1,.)\cdots \phi(x_k,.)y\k$$
is to be understood as a function on the domain $\Md$.

Consider the image
$$\Gdk := \ed(X^{\otimes k}\otimes Y\otimes \Fdd)$$
and the union over all $k$
$$\Gd := \bigcup_{k\in\bN}\Gdk\,.$$
The construction of the topology proceeds in two steps:
first we show that $W$ can be embedded into
$\Gd$; then
a topology is given to $\Gd$ and thus also to $W$.

The space $W$ can be embedded into $\Gd$ as follows:
for any $k$-tuple $n_1,\ldots,n_k\in\bZ$ one
defines functionals $\mu_{n_1,\ldots,n_k}\in\Fdd$ by
\begin{eqnarray*}
\lefteqn{\mu_{n_1,\ldots,n_k}
(\b\wt,\phi(v_1,.)\cdots \phi(v_k,.)w\k)} \\ \nonumber
&&=
\frac{1}{2\pi\i}\oint_{|z_1|=r_1}\cdots
\frac{1}{2\pi\i}\oint_{|z_k|=r_k}z_1^{n_1}\cdots
z_k^{n_k}\,\b v',\phi(v_1,z_1)\cdots \phi(v_k,z_k)w\k\,
\d z_1\cdots\d z_k \\
&&=
\b\wt,(v_1)_{n_1}\cdots (v_k)_{n_k}w\k\,,
\end{eqnarray*}
where $r > r_1>\cdots >r_k > 0$.
Note that the inner product provides an isomorphism
between $W$ and a subspace of $\overline{\rm Hom}(W,\bC)$.
As explained in section 1, $W$ is spanned by vectors of the form
$$w = (x_1)_{n_1}\cdots (x_k)_{n_k} y\qquad (k\in\bN)$$
where $x_1,\ldots x_k\in X$, $n_1,\ldots, n_k\in\bZ_+$ and $y\in Y$.

The value of the inner product $\b\;\:, w\k$ coincides with the value of the
functional
$$\ed(x_1\otimes\cdots\otimes x_k\otimes y\otimes\mu_{n_1,\ldots,n_k})$$
on any vector $\tilde w \in W$. Indeed,
\begin{eqnarray*}
\lefteqn{
\ed(x_1\otimes\cdots\otimes x_k\otimes y\otimes\mu_{n_1,\ldots,n_k})(\wt)} \\
&&= \mu_{n_1,\ldots,n_k}(\b \wt, \phi(x_1,.)\cdots \phi(x_k,.)y\k) \\
&&= \b \wt,(x_1)_{n_1}\cdots (x_k)_{n_k}y\k \\
&&= \b \wt,w\k\,.
\end{eqnarray*}
Thus, $w$ can be identified with the vector $\ed(x_1\otimes\cdots\otimes x_k\otimes y\otimes\mu_{n_1,\ldots,n_k})$
in $\Gdk\subset \Gd$. This defines our embedding of $W$ into $\Gd$.

Next one constructs a canonical embedding of
$\Gdk$ into $G^{\D}_{k+1}$: define the linear map
$$\gak: F^{\D}_{k+1}\to\Fd$$
by
$$\gak(\b\wt,\phi(v_1,.)\cdots\phi(v_{k+1},.)w\k)
= \b (v_1)_{-1}^* \wt,\phi(v_2,.)\cdots\phi(v_{k+1},.)w\k\,.$$
$(v_1)_{-1}^*$ is the adjoint of the $-1$st mode of $\phi(v_1,z_1)$.
For arbitrary $\wt\in W$, we have
\begin{eqnarray*}
\lefteqn{
\ed(x_1\otimes\cdots\otimes x_k\otimes y\otimes\mu)(\wt)} \\
&&= \mu(\b \wt, \phi(x_1,.)\cdots \phi(x_k,.)y\k) \\
&&= \mu(\gak(\b \wt,\phi(\Omega,.)\phi(x_1,.)\cdots \phi(x_k,.)y\k)) \\
&&= (\gak^*(\mu))(\b \wt,\phi(\Omega,.)\phi(x_1,.)\cdots \phi(x_k,.)y\k) \\
&&= e_{k+1}^{\D}(\Omega\otimes x_1\otimes\cdots\otimes x_k\otimes
y\otimes\gak^*(\mu))(\wt)\,,
\end{eqnarray*}
where the adjoint
$$\gak^*:\Fdd\to F^{\D*}_{k+1}$$
has been used. This shows that $\Gdk\subset G^{\D}_{k+1}$,
and that the union $\Gd$ of all
such spaces is a  vector space.

How can $\Gd$ be made topological?
Both $X$ and $Y$ are finite-dimensional and carry a unique Banach space structure.
So does the tensor product $X^{\otimes k}\otimes Y$. $\Fdd$ has the strong topology, and
we equip $X^{\otimes k}\otimes Y\otimes\Fdd$ with the projective tensor product
topology. $\Gdk$ is the image of $X^{\otimes k}\otimes Y\otimes\Fdd$ under the
linear and surjective map $\ed$, and is given the {\it final
(identification) topology}. It can
be shown then that for any $k\in\bN$, $\Gdk$
is a topological subspace of $\Gd_{k+1}$. We have an increasing sequence of
locally convex spaces whose union yields the vector space $\Gd$.
The topology on $\Gd$ is defined as the strict inductive limit
determined by this sequence.

As a subspace, $W$ inherits a locally convex and Hausdorff topology from $\Gd$;
we denote it by $\THud$.

The proofs are analogous to those in
section 1 and 3 of \cite{Huang} except for the following
replacements:
$\tG$ becomes $W$, i.e.\ the value
$\b\lambda,w\k$ of a functional $\lambda\in\tG$ on a vector
$w\in W$ is replaced by the inner product $\b\wt,w\k$
between a vector $\wt\in W$ and $w$. Instead of the dual space $\tG^*$
we use the space $\overline{\rm Hom}(W,\bC)$ of antilinear
functionals on $W$, equipped with the weak topology.
The function spaces in \cite{Huang} correspond to $\Fd$ with $\D$ the open unit
disk in $\bC$.

\section{Properties of the Topologies}
\label{Properties_of_the_Topologies}
The proofs in section \ref{Properties_of_the_Topologies} and
\ref{Comparison_of_the_Topologies}
are again formulated for a general finitely-generated $V$-module $W$.

\subsection{Nuclearity}
\label{Nuclearity}
We show below that $\THud$ is nuclear, and that $\TGG$ is nuclear if it is
Hausdorff.
Gaberdiel's and Goddard's space of states results from dividing $W$ by the ``null states'' with respect to $\TGG$, which renders it Hausdorff and nuclear.

 In the proof the following properties of nuclear spaces are used:
\begin{enumerate}
\item A linear subspace of a nuclear space is nuclear.
\item \label{quotient}The quotient of a nuclear space modulo a closed linear subspace is
nuclear.
\item\label{projective} A projective limit of nuclear spaces is
nuclear if it is Hausdorff.
\item A countable inductive limit of nuclear spaces is
nuclear.\label{inductive}
\item The projective tensor product of two nuclear
spaces is  nuclear. \label{tensorproduct}
\item A Fr\'{e}chet space is nuclear if and only if its strong
dual is nuclear. \label{Frechetdual}
(A topological vector space is called a Fr\'{e}chet space if
it is complete, metrizable, locally convex and Hausdorff.)
\end{enumerate}
For detailed definitions and proofs see, for instance, \cite{Treves},
chap.\ 50. The following theorem provides an alternative characterization
for locally convex metrizable spaces:
\begin{enumerate}
\item[7.] A locally convex space is metrizable iff
its topology can be described by a countable family of seminorms.
\end{enumerate}

Given some open subset $D$ of $\bC^n$, $n\in\bN$,
the space $H(D)$ of holomorphic functions on it is nuclear (\cite{Treves}, chap.\ 51).
Accordingly, $H(\Mo)$ and $H(\Md)$ are nuclear spaces, and the same
holds true for the subspaces
$\Fo$, $\Fd$ and $\hFd$.
$\TGG$ is a projective limit of the spaces $\Fo$, and hence
nuclear if it is Hausdorff (\pt{projective}).
If $W$ is divided by all null states, the projective limit becomes Hausdorff and therefore nuclear (\pt{projective}).

Consider now Huang's topology:
Clearly, $\Fd$ is an example of a Fr\'{e}chet space.
By \pt{Frechetdual}\ the strong dual $\Fdd$ of $\Fd$ is
nuclear. Note that this conclusion cannot be made for $\hFdd$, since
$\hFd$ may not be complete and hence not a Fr\'{e}chet space.
It follows from the definition that
$$\Gdk \cong (X^{\otimes k}\otimes Y\otimes\Fdd)/(\ed)^{-1}(0)\,,$$
where $\cong$ denotes a linear and topological isomorphism.
The finite-dimensional space $X^{\otimes k}\otimes Y$ is nuclear and
according to \pt{tensorproduct}\ the tensor product with $\Fdd$
is nuclear as well. $\ed$ is continuous (Proposition 1.5 in \cite{Huang})
and $(\ed)^{-1}(0)$ closed, so
\pt{quotient}\ tells us that $\Gdk$ has the nuclear property.
By \pt{inductive}, the latter is preserved under the inductive limit
$$\Gd = \bigcup_{k\in\bN}\Gdk\,,$$
and $W$, as a subspace of $\Gd$, must again be nuclear.
\subsection{Dependence on $\O$ and $\D$}
\label{Dependence_on_O_and_D}
In the case of Gaberdiel's and Goddard's topologies, it is immediate
from the definition that
$$\mbox{$\TGGp$ is coarser than $\TGG$ when $\O'\subset\O$.}$$
Due to their definition by functionals, Huang's topologies behave in the opposite way:
$$\mbox{$\THuc$ is finer than $\THud$ for $\C\subset\D$,}$$
This can be seen as follows:

\noindent\pf
Suppose that $\C\subset\D\subset\bC$ where $\C$ and $\D$ are open disks
centered at 0. Consider the map from $\Fd$ to $\Fc$ given by restriction
to $\Mc$: it is linear, surjective, and injective, since both pre-image and
image are restrictions of a single {\it meromorphic}
function on the domain $\MC = M^k$. $\Fc$ and $\Fd$ can be
identified as vector  spaces, but the topology on $\Fc$ is weaker. Therefore,
its dual space $\Fcd$ is a subspace of $\Fdd$. Both dual spaces carry the strong
topology: the topology of uniform convergence on weakly bounded\footnote{Bounded = weakly
bounded in locally convex Hausdorff spaces (see \cite{Narici}, (9.7.6)).} subsets.
Neighbourhood bases at 0 are given by the polar sets
$$\Bpd = \{\mu\in\Fdd\,|\,\sup_{f\in B} |\mu(f)| \le 1\}\quad\mbox{for $B$ bounded in $\Fd$,}$$
and
$$\Bpc = \{\mu\in\Fcd\,|\,\sup_{f\in B} |\mu(f)| \le 1\}\quad\mbox{for $B$ bounded in $\Fc$}$$
respectively.
The topology induced on $\Fcd$ by $\Fdd$ has the
base
$$\tBpc = \{\mu\in \Fcd\,|\,\sup_{f\in B} |\mu(f)| \le 1\}\quad\mbox{for $B$ bounded in $\Fd$.}$$
A set $B$ is bounded in $\Fd$ iff it is bounded w.r.t.\ each
seminorm in $\Fd$. Hence it is also bounded in $\Fc$, and $\tBpc = \Bpc$.
Thus we see that the topology on $\Fcd$ is finer than that induced
by $\Fdd$.

Furthermore, as topological vector spaces,
$$\Gdk \cong
(X^{\otimes k}\otimes Y\otimes\Fdd)/(\ed)^{-1}(0)\,,$$
and
$$\Gck \cong
(X^{\otimes k}\otimes Y\otimes\Fcd)/(\ec)^{-1}(0)\,,$$
Since $\ec$ is simply the restriction of $\ed$ to
$X^{\otimes k}\otimes Y\otimes\Fcd$, $\Gck$ is a subspace of $\Gdk$.
By definition, the topology on $X^{\otimes k}\otimes Y\otimes\Fdd$
is the projective tensor product of
$X^{\otimes k}\otimes Y$ and $\Fdd$.
A neighbourhood base at 0
of the space $X^{\otimes k}\otimes Y \otimes \Fdd$
is constituted by the sets
$$\conv(U\otimes N)$$
where $U$ and $N$ are neighbourhoods in
$X^{\otimes k}\otimes Y$ and $\Fdd$ respectively.
The set $U\otimes N$ consists
of all $u\otimes\mu$, $u\in U$, $\mu\in N$, and
conv stands for the convex hull. We have
\begin{eqnarray*}
\lefteqn{\conv(U\otimes N)\cap(X^{\otimes k}\otimes Y\otimes\Fcd)} \\
&\supset&\conv((U\otimes N)\cap(X^{\otimes k}\otimes Y\otimes\Fcd)) \\
&\supset&\conv(U\otimes (N\cap\Fcd))\,,
\end{eqnarray*}
and $N\cap\Fcd$ is a neighbourhood of 0 in $\Fcd$.
This implies that the topology of
$X^{\otimes k}\otimes Y\otimes\Fcd$ is finer than that induced
on it by $X^{\otimes k}\otimes Y\otimes\Fdd$. Therefore, the topology of
$\Gck$ is finer than that induced by $\Gdk$.
The same holds true
for the inductive limits $\Gc$ and $\Gd$, and we conclude that $\THuc$ is finer than
$\THud$.
\qed
\section{Comparison of the Topologies}
\label{Comparison_of_the_Topologies}
We would like to show that $\THud$ is finer than $\TGG$
for suitable choices of $\cal D$ and $\cal O$. For that purpose it
suffices to prove that each seminorm of Gaberdiel's \& Goddard's topology
is continuous in Huang's topology. In the notation of sec.\ 3, this means that for each
$l\in\bN$, $I\in\{1,\ldots,n\}\times\{1,\ldots,d\}^l$ and compact subset $K\subset\Mo$,
the seminorm
$$\|.\|_{I,K} := \|\go(x_I\otimes .)\|_K$$
is continuous in $\THud$. Let us therefore consider $I$ and $K$ to be fixed.
We have to show that for any net $\{w_s\}_{s\in S}$ ($S$ an index set)
that converges to 0 in $\THud$, the net
\begin{eqnarray}
\|w_s\|_{I,K} & = & \|\b y_{i_0},\phi(x_{i_1},.)\cdots
\phi(x_{i_l},.)w_s\k\|_K \nonumber \\
& = & \sup_{\v{\S\zeta}\in K} |\b y_{i_0},\phi(x_{i_1},\zeta_1)\cdots
\phi(x_{i_l},\zeta_l)w_s\k| \label{fixedIK}
\end{eqnarray}
goes to 0 as well. To simplify notation we drop the index $i$
and write $y, x_1,\ldots,x_l$ from now on.

The proof proceeds in three steps: We specify a neighbourhood base at 0
for $\THud$ and express the convergence of $\{w_s\}_{s\in S}$ in terms
of it. To apply this convergence property, we need to cast the correlator
$$\b y_{i_0},\phi(x_{i_1},\zeta_1)\cdots
\phi(x_{i_l},\zeta_l)w_s\k$$
into a different form. Eq.\ \eq{result} below provides the desired
reordering, and is proved by using the Laurent expansion of
correlation functions. This equality is also essential for the proof in sec.\ \ref{Continuity_of_Conformal_Blocks}.
The third step consists in choosing a neighbourhood at 0 of $\THud$ such that \eq{fixedIK}
becomes smaller than a given $\epsilon$.
\subsection{Convergence in Huang's Topology}
\label{Convergence_in_Huangs_Topology}
Let us recall what spaces were involved in the construction of
Huang's topology:
$X^{\otimes k}\otimes Y$ is of finite dimension $nd^k$ and has a
norm topology. All norms on $X^{\otimes k}\otimes Y$ are
equivalent, so we can take it to be the 1-norm w.r.t.\ some basis
(i.e.\ the sum of the absolute values of the coefficients in this basis).
Let $U_{\delta}(0)$ denote the associated ball of radius $\delta > 0$
around 0. $\Fdd$ carries the strong topology, and a base
for the neighbourhoods of $0$ in $\Fdd$ is given by the polars
$$\Bp = \{\mu\in\Fdd\,|\,\sup_{f\in B} |\mu(f)|
\le 1\}$$ where $B$ is bounded in $\Fd$. As already mentioned in
sec.\ \ref{Dependence_on_O_and_D}, a neighbourhood base at 0 for
$X^{\otimes k}\otimes Y\otimes \Fdd$ is provided
by the sets $\conv(U\otimes N),$
where $U$ and $N$ are neighbourhoods in $X^{\otimes k}\otimes Y$
and $\Fdd$ respectively. Clearly, the sets
$$\conv(U_{\delta}(0)\otimes\Bp)\qquad\mbox{($\delta > 0$, $B$ bounded in
$\Fd$)}$$
form an equivalent base.
The space $\Gdk$ is the image of $X^{\otimes k}\otimes Y\otimes \Fdd$
under the map $\ed$ and carries the associated {\it final} or {\it identification}
topology. Therefore, the sets
\begin{equation}
\label{baseGdk}
\ed(\conv(U_{\delta}(0)\otimes\Bp))
\end{equation}
provide us with a neighbourhood base at 0 for $\Gdk$. The space
$$\Gd = \bigcup_{k\in\bN}\Gdk$$
is the strict inductive limit of the spaces $\Gdk$, and
induces the topology $\THud$ on its (embedded) subspace $W$.
A base at 0 for $\Gd$ is constituted
by the sets of the form
\begin{equation}
\label{baseGd}
\conv\left(\bigcup_{k\in\bN}{\cal U}_k\right)\,,
\end{equation}
where each ${\cal U}_k$ is a neighbourhood of 0 in $\Gdk$
(see \cite{Narici}, p.287, sec.\ 12.1).
Combining \eq{baseGdk} and \eq{baseGd}, we see that the sets
$$W\:\cap\:\conv\!\!\left(\bigcup_{k\in\bN}\ed(\conv(U_{\delta_k}(0)\otimes\Bp_k))\right)$$
give a base at 0 for Huang's topology\footnote
{It is understood that
$U_{\delta_k}(0)$ and $\Bp_k$ belong to the spaces $X^{\otimes k}\otimes Y$
and $\Fdd$ respectively.}. Since $\ed$ is linear, the latter simplifies
to
\begin{equation}
\label{baseW}
W\:\cap\: \conv\!\!\left(\bigcup_{k\in\bN}\ed(U_{\delta_k}(0)\otimes\Bp_k)\right)\,.
\end{equation}
Note that in writing so we have
identified $W$ with its image under the embedding in $\Gd\subset\overline{\rm Hom}(W,\bC)$.

Consider now the net $\{w_s\}_{s\in S}$ which converges to 0 in the topology
$\THud$ on $W$. Given a sequence of pairs $(\delta_k,B_k)$, $k\in\bN$,
there is an index $s_0$ such that for each $s\ge s_0$, $w_s$ can
be expressed as a finite sum
\begin{equation}
\label{finite_sum}
w_s = \sum_i a_i\,\edi(u_i\otimes\mu_i)
\end{equation}
with
$$u_i\in U_{\delta_{k_i}}(0)\,,\quad\mu_i\in\Bp_{k_i}\,,\qquad k_i\in\bN\,,$$
and coefficients obeying
$$\sum_i |a_i| \le 1\,.$$
To simplify notation we write the right-hand side of \eq{finite_sum} without
index $s$.
\subsection*{Laurent Expansion}
We want to give an upper estimate for expression \eq{fixedIK} when $s\ge s_0$.
Let us first consider
the case when
the sum \eq{finite_sum} consists of only one term, i.e.
$$w_s = \ed(u\otimes\mu)\,,\qquad\mbox{$u\in U_{\delta_k}(0)$, $\mu\in\Bp_k$}$$
for some $s\ge s_0$ and $k\in\bN$.
In the following the index $k$ is fixed, so we will omit it from
$U_{\delta}(0)$ and $B$.

The correlator in \eq{fixedIK} can now be written as
\begin{equation}
\label{correlatorws}
\b y_0,\phi(x_1,\zeta_1)\cdots
\phi(x_l,\zeta_l)w_s\k
= \b y_0,\phi(x_1,\zeta_1)\cdots
\phi(x_l,\zeta_l)\ed(u\otimes\mu)\k\,.
\end{equation}
In \eq{correlatorws}
we would like to apply the definition of $\ed$
and make the functional $\mu$ appear explicitly (see eq.\ \eq{ek}).
The operators $\phi(x_1,\zeta_1)\cdots \phi(x_l,\zeta_l)$ prevent us from
doing so and should be removed somehow. Given a $\v{\zeta} =
(\zeta_1,\ldots,\zeta_l)\in\Mo$ such that
$$|\zeta_1|>\ldots>|\zeta_l|>0\,,$$
one can expand the correlation function in its natural power
series
\begin{eqnarray}
\lefteqn{\sum_{m\in\bZ^l}\b y,(x_1)_{m_1}\cdots
(x_l)_{m_l}\ed(u\otimes\mu)\k\,\zeta_1^{-m_1-1}\cdots\zeta_l^{-m_l-1}}
\nonumber
\\
& = &
\sum_{m\in\bZ^l}\b (x_l)^*_{m_l}\cdots
(x_1)^*_{m_1} y,\ed(u\otimes\mu)\k\,\zeta_1^{-m_1-1}\cdots\zeta_l^{-m_l-1} \nonumber
\\
& = & \label{xtotheleft}
\sum_{m\in\bZ^l}\ed(u\otimes\mu)((x_l)^*_{m_l}\cdots (x_1)^*_{m_1} y)\,\zeta_1^{-m_1-1}\cdots\zeta_l^{-m_l-1}
\end{eqnarray}
Next we specify a basis
$$u^j = u^j_1\otimes\cdots\otimes u^j_k\otimes u^j_{k+1}\,,$$
$$\mbox{$u^j_1,\ldots,u^j_k\in X$, $u^j_{k+1}\in Y$,}\quad j = 1,\ldots, nd^k\,,$$
for
$X^{\otimes k}\otimes Y$,
and choose the associated 1-norm to be the norm on $X^{\otimes k}\otimes Y$.
Then, each $u\in U_{\delta}(0)\subset X^{\otimes k}\otimes Y$ is a linear
combination
$$u = \sum_{j=1}^{nd^k} b_j\,u^j\,,\qquad |b_j|\le\delta\,,$$
and after applying the definition of $\ed$, the power series \eq{xtotheleft}
becomes
\begin{eqnarray}
\lefteqn{\sum_{j=1}^{nd^k} b_j\sum_{m\in\bZ^l}
\ed(u^j\otimes\mu)((x_l)^*_{m_l}\cdots (x_1)^*_{m_1} y)
\,\zeta_1^{-m_1-1}\cdots\zeta_l^{-m_l-1}}
\nonumber \\
& = &
\sum_{j=1}^{nd^k} b_j\sum_{m\in\bZ^l}\mu(\b (x_l)^*_{m_l}\cdots
(x_1)^*_{m_1} y,\phi(u^j_1,.)\cdots
\phi(u^j_k,.)u^j_{k+1}\k) \nonumber \\
& & \times\,\zeta_1^{-m_1-1}\cdots\zeta_l^{-m_l-1}\,.
\label{muLaurent}
\end{eqnarray}
Each term in the sum over $j$ looks like the functional $\mu$ applied
to
\begin{eqnarray}
\lefteqn{\sum_{m\in\bZ^l}\b (x_l)^*_{m_l}\cdots
(x_1)^*_{m_1} y,\phi(u^j_1,.)\cdots
\phi(u^j_k,.)u^j_{k+1}\k
\,\zeta_1^{-m_1-1}\cdots\zeta_l^{-m_l-1}} \nonumber \\
&& =
\sum_{m\in\bZ^l}\b y,(x_1)_{m_1}\cdots
(x_l)_{m_l}\phi(u^j_1,.)\cdots
\phi(u^j_k,.)u^j_{k+1}\k
\,\zeta_1^{-m_1-1}\cdots\zeta_l^{-m_l-1}\,.
\label{Hartogs}
\end{eqnarray}
Note that \eq{Hartogs} is a
Hartogs expansion in $\v{\zeta}$ of\footnote{Be reminded that
dots represent variables of the function, whereas
$\zeta_1$ to $\zeta_l$ are fixed.}
\begin{equation}
\label{limit}
f_j = \b y, \phi(x_1,\zeta_1)\cdots
\phi(x_l,\zeta_l) \phi(u^j_1,.)\cdots \phi(u^j_k,.)\,u^j_{k+1}\k\,,
\end{equation}
provided that
$$\sup_{z\in\D} |z| < |\zeta_l|\,.$$
The partial sums of \eq{Hartogs} take their values in the dense subspace $\hFd$ of $\Fd$.
It is a sequence of functions in $\hFd$,
but in general not convergent to a function of $\hFd$.
At this point it
becomes important that in the construction of $\THud$
we have used the completion $\Fd$ instead of $\hFd$.
A theorem of complex analysis states that
the Hartogs series \eq{Hartogs} converges compactly
to $f_j$. As a result, $f_j$ is contained in the completion $\Fd$, and with $\mu$ being
an element of $\Fdd$ the infinite sums in \eq{muLaurent} can be written as
\begin{eqnarray*}
\lefteqn{\sum_{m\in\bZ^l}\mu(\b y,(x_1)_{m_1}\cdots(x_l)_{m_l}
\phi(u^j_1,.)\cdots \phi(u^j_k,.)u^j_{k+1}\k)\,\zeta_1^{-m_1-1}\cdots\zeta_l^{-m_l-1}}
\\
&&= \mu(\b y, \phi(x_1,\zeta_1)\cdots
\phi(x_l,\zeta_l) \phi(u^j_1,.)\cdots \phi(u^j_k,.)\,u^j_{k+1}\k)\,.
\end{eqnarray*}
Recalling our starting point (eq.\ \eq{correlatorws}) we get
\begin{eqnarray}
\lefteqn{\b y, \phi(x_1,\zeta_1)\cdots
\phi(x_l,\zeta_l)\,\ed(u\otimes\mu)\k} \nonumber \\
&& = \sum_{j=1}^{nd^k} b_j\,\mu(\b y, \phi(x_1,\zeta_1)\cdots
\phi(x_l,\zeta_l) \phi(u^j_1,.)\cdots \phi(u^j_k,.)\,u^j_{k+1}\k)\,.
\label{result}
\end{eqnarray}
Let us recollect what assumptions were needed in order to arrive at
the relation \eq{result}:
The values of $\zeta_1,\ldots,\zeta_l$ are taken from an open subset
$\O$ of $\bC$ and \eq{limit}
makes only sense as a function on $\Md$ if $\O$ and $\D$ do not
overlap. Furthermore, the Hartogs expansion \eq{Hartogs} requires
that
\begin{equation}
\label{order}
|\zeta_1|>\ldots>|\zeta_l|>\sup_{z\in\D}|z|\,.
\end{equation}
Eq.\ \eq{result} continues to hold for arbitrary orderings of
$\zeta_1,\ldots,\zeta_l\in\O$ provided
$$\inf_{\zeta\in\O} |\zeta| > \sup_{z\in\D} |z| = r\,.$$
For $i\neq j$, $\zeta_i\neq \zeta_j$, but what about values where
$|\zeta_i| = |\zeta_j|$?
Let $\v{\zeta}_0 = (\zeta_{1,0},\ldots \zeta_{l,0})$ be a point where at least
two radii coincide.
Clearly, there is a sequence $\v{\zeta}_n$ in a region
of the type \eq{order} such that
$$\lim_{n\to\infty}\v{\zeta}_n = \v{\zeta}_0\,.$$
The corresponding sequence of functions
$$\b y, \phi(x_1,\zeta_{1,n})\cdots
\phi(x_l,\zeta_{l,n}) \phi(u^j_1,.)\cdots \phi(u^j_k,.)\,u^j_{k+1}\k$$
converges compactly to
$$\b y, \phi(x_1,\zeta_{1,0})\cdots
\phi(x_l,\zeta_{l,0}) \phi(u^j_1,.)\cdots \phi(u^j_k,.)\,u^j_{k+1}\k\,.$$
By continuity of $\mu$ it follows that equation \eq{result} is valid for
$\v{\zeta}_0$ and thus for arbitrary values of $\v{\zeta}\in\Mo$.
\subsection*{Choice of Neighbourhood}
Given $\epsilon > 0$,
we seek neighbourhoods $U_{\delta}(0)$ and
$\Bp$ such that
$$\sup_{\v{\S\zeta}\in K}|\b y,\phi(x_1,\zeta_1)\cdots
\phi(x_l,\zeta_l)\,\ed(u\otimes\mu)\k| \le\epsilon$$
if $u\in U_{\delta}(0)$ and $\mu\in\Bp$.

For each point $\vz = (\zeta_1,\ldots,\zeta_l)\in K$ there is
a correlation function
$$f_j = \b y, \phi(x_1,\zeta_1)\cdots
\phi(x_l,\zeta_l) \phi(u^j_1,.)\cdots \phi(u^j_k,.)\,u^j_{k+1}\k\,.$$
Let $B_j$ denote the set of these functions.
When regarded
as a function of $l+k$ variables,
$$\tfj = \b y, \phi(x_1,.)\cdots
\phi(x_l,.) \phi(u^j_1,.)\cdots \phi(u^j_k,.)\,u^j_{k+1}\k$$
is holomorphic on $\Mo\times\Md$, and for any compact
subset $K'$ of $\Md$
$$\sup_{(\v{\S\zeta},\v{\S z})\in K\times K'}
|\tfj(\vz,\vzd)|\,<\,\infty\,.$$
This means that $B_j$ is bounded in $\Fd$, and the same
holds true for the union over $j$
$$B = \bigcup_{j=1}^{nd^k} B_j\,.$$
Then, if $\mu\in\Bp$ and $u\in U_{\delta}(0)$, $\delta = \epsilon/(nd^k)$,
eq.\ \eq{result} implies that
\begin{eqnarray*}
\lefteqn{|\b y,\phi(x_{i_1},\zeta_1)\cdots
\phi(x_{i_l},\zeta_l)\,\ed(u\otimes\mu)\k|} \\
&& \le\,
\sum_{j=1}^{nd^k} |b_j|\,|\mu(\b y, \phi(x_1,\zeta_1)\cdots
\phi(x_l,\zeta_l) \phi(u^j_1,.)\cdots \phi(u^j_k,.)\,u^j_{k+1}\k)| \\
&& \le nd^k\delta\le\,\epsilon\qquad\forall\,\v{\zeta}\in K\,,
\end{eqnarray*}
as required.
In general, $w_s$ is a finite sum
of the type \eq{finite_sum} for $s\ge s_0$, and one has to
consider the expression
$$|\,\b y,\phi(x_1,\zeta_1)\cdots
\phi(x_l,\zeta_l)\,\sum_i a_i\,\edi(u_i\otimes\mu_i)\k\,|\,.$$
Using linearity and
$$\sum_i |a_i| \le 1$$
we can repeat the same arguments to obtain
$$\sup_{\v{\S\zeta}\in K} |\b y,\phi(x_1,\zeta_1)\cdots
\phi(x_l,\zeta_l)w_s\k| \le \epsilon$$
for $s\ge s_0$.
Therefore the seminorms $\|.\|_{I,K}$ are continuous in the
topology $\THud$. The proof was based on the validity of \eq{result}, i.e.\ we need that
$$\inf_{\zeta\in\O} |\zeta| > \sup_{z\in\D} |z| = r\,,$$
and in that case $\THud$ is finer than $\TGG$. \qed
\section{Continuity of Conformal Blocks}
\label{Continuity_of_Conformal_Blocks}
Associated to a choice of $m$ points $z_1,\ldots,z_n\in\bC$,
we define a conformal $m$-point block as the linear
functional
\begin{equation}
\label{block}
\renewcommand{\arraystretch}{1.5}
\begin{array}{llll}
C_m(z_1,\ldots,z_m): & V^{\otimes m} & \to & \bC\,, \\
& v_1\otimes\cdots\otimes v_m & \mapsto &
\b\Omega,\phi(v_1,z_1)\cdots\phi(v_m,z_m)\Omega\k\,.
\end{array}
\end{equation}
Note that we are now dealing with matrix elements of
the vertex algebra $V$. Physically speaking, these are
conformal blocks whose insertions are in the bosonic vacuum
sector $V$; we do not consider conformal blocks of other
sectors, since we have not introduced general intertwining
operators.

In an arbitrary topology on $V$, the functionals \eq{block}
need not be continuous. We present a proof that
two-point blocks are continuous in Huang's topology
$\THud$, provided $\D$ is small enough.
The method can be generalized to arbitrary
$m$-point blocks in principle,
though it becomes rather unwieldy for $m>2$.

For fixed points $z, \zt\in\bC$, the two-point block
$$
\renewcommand{\arraystretch}{1.5}
\begin{array}{llll}
C_2(z,\zt): & V\otimes V & \to & \bC\,, \\
& v\otimes \vt & \mapsto &
\b\Omega,\phi(\vt,\zt)\phi(v,z)\Omega\k\,.
\end{array}
$$
is continuous on $V\otimes V$ iff it is
continuous as a bilinear map from $V\times V$
into the complex numbers.
A net $\{(\vt_s,v_s)\}_{s\in S}$
converges to 0 in $V\times V$ iff both
$\{\vt_s\}_{s\in S}$ and $\{v_s\}_{s\in S}$
converge to 0 in $V$.
Given such nets we want to demonstrate that
\begin{equation}
\label{correlatorvs}
|\b\Omega,\phi(\vt_s,\zt)\phi(v_s,z)\Omega\k|
\end{equation}
goes to zero.

The proof is similar to the one of sec.\ \ref{Comparison_of_the_Topologies}:
We express the convergence of $v_s$ and $\vt_s$ in terms of
neighbourhoods at 0, and manipulate expression \eq{correlatorvs}
such that eq.\ \eq{result} can be applied. Then we choose suitable neighbourhoods
to make the value of \eq{correlatorvs} smaller than $\epsilon$.

Repeating arguments of sec.\ \ref{Convergence_in_Huangs_Topology}
we see that given a sequence of $\delta_k > 0$ and bounded sets
$B_k$, $\Bt_k$ in $\Fd$, there is an index $s_0$
such that for each $s\ge s_0$, $v_s$ and $\vt_s$ are finite sums of the type
\begin{equation}
\label{convex_hull}
v_s = \sum_i a_i\,\edi(u_i\otimes\mu_i)\,,\quad
\vt_s = \sum_i \at_i\,\edi(\ut_i\otimes\mut_i)\,,
\end{equation}
with
$$\mbox{$u_i, \ut_i\in U_{\delta_{k_i}}(0)\,$,\quad $\mu_i\in\Bp$, $\mut_i\in\Btp$},\qquad k_i\in\bN\,,$$
and coefficients obeying
$$\sum_i |a_i| \le 1\,,\quad\sum_i |\at_i| \le 1\,.$$
Suppose for the moment that for an $s\ge s_0$ each of the two sums contains
only one term, that is
$$v_s = \ed(u\otimes\mu)\,,\quad\vt_s = \ed(\ut\otimes\mut)\,,$$
and
$$\mbox{$u, \ut\in U_{\delta_k}(0)\,,$\quad $\mu\in\Bp_k, \mut\in\Btp_k$}$$
for some $k\in\bN$. Below the index $k$ is omitted from $U_{\delta}(0)$,
$B$ and $\Bt$.
Again, we write $u$ and $\ut$ as linear combinations
of orthonormal basis vectors:
$$u = \sum_{j=1}^{nd^k} b_j\,u^j\,,\quad \ut = \sum_{j'=1}^{nd^k} \bt_{j'}\,\ut^{j'}, $$
where
$$|b_j|\le\delta\,,\quad u^j = u^j_1\otimes\cdots\otimes u^j_k\otimes u^j_{k+1}\,,$$
$$\mbox{$u^j_1,\ldots,u^j_k\in X$, $u^j_{k+1}\in Y$,}\quad j = 1,\ldots, nd^k\,,$$
and
$$|\bt_{j'}|\le\delta\,,\quad \ut^{j'} = \ut^{j'}_1\otimes\cdots\otimes \ut^{j'}_k\otimes \ut^{j'}_{k+1}\,,$$
$$\mbox{$\ut^{j'}_1,\ldots,\ut^{j'}_k\in X$, $\ut^{j'}_{k+1}\in Y$,}\quad j' = 1,\ldots, nd^k\,.$$
We are now ready to express the two-point correlator
$\b\Omega,\phi(\vt_s,\zt)\phi(v_s,z)\Omega\k$
in terms of the defining maps of Huang's topology.
The calculation employs translation invariance (t),
locality (l), equation \eq{result} and the state-operator
correspondence (s). Within correlation functions
$\phi(v,0)$ stands for the zero limit in the complex
variable.
\begin{eqnarray}
\lefteqn{\b\Omega,\phi(\vt_s,\zt)\phi(v_s,z)\Omega\k} \nonumber \\
&\stackrel{t}{=}& \b\Omega,\phi(\vt_s,\zt-z)\phi(v_s,0)\Omega\k \nonumber \\
&=& \b\Omega,\phi(\vt_s,\zt-z)\ed(u\otimes\mu)\k \nonumber \\
&\stackrel{\eq{result}}{=}& \sum_{j=1}^{nd^k} b_j\,\mu(\b\Omega,\phi(\vt_s,\zt-z)
\phi(u^j_1,.)\cdots
\phi(u^j_k,.)\,u^j_{k+1}\k) \nonumber \\
&\stackrel{s}{=}& \sum_{j=1}^{nd^k} b_j\,\mu(\b\Omega,\phi(\vt_s,\zt-z)\phi(u^j_1,.)\cdots
\phi(u^j_k,.)\phi(u^j_{k+1},0)\Omega\k) \nonumber \\
&\stackrel{l,t}{=}& \sum_{j=1}^{nd^k} b_j\,\mu(\b\Omega,\phi(u^j_1,.-\zt+z)\cdots
\phi(u^j_k,.-\zt+z)\phi(u^j_{k+1},-\zt+z)\phi(\vt_s,0)\Omega\k) \nonumber \\
&\stackrel{\eq{result}}{=}& \sum_{j=1}^{nd^k} b_j\sum_{j'=1}^{nd^k} \bt_{j'}\,
\mu(\mut(\b\Omega,\phi(u^j_1,.-\zt+z)\cdots
\phi(u^j_k,.-\zt+z)\phi(u^j_{k+1},-\zt+z) \nonumber \\
& &
\hspace{2.7cm}\times\,\phi(\ut^{j'}_1,\dt)\cdots\phi(\ut^{j'}_k,\dt)\ut^{j'}_{k+1}\k))
\nonumber \\
&\stackrel{t}{=}& \sum_{j=1}^{nd^k} b_j\sum_{j'=1}^{nd^k} \bt_{j'}\,
\mu(\mut(\b\Omega,\phi(u^j_1,z+.)\cdots
\phi(u^j_k,z+.)\phi(u^j_{k+1},z) \nonumber \\
& &
\hspace{2.7cm}\times\,\phi(\ut^{j'}_1,\zt+\dt)\cdots\phi(\ut^{j'}_k,\zt+\dt)\phi(\ut^{j'}_{k+1},\zt)\Omega\k))\,.
 \label{expanded}
\end{eqnarray}
The notation should be understood as follows: $\mut$ acts on
$$
\b\Omega,\phi(u^j_1,z+.)\cdots
\phi(u^j_k,z+.)\phi(u^j_{k+1},z)\phi(\ut^{j'}_1,\zt+\dt)\cdots\phi(\ut^{j'}_k,\zt+\dt)\phi(\ut^{j'}_{k+1},\zt)\Omega\k
$$
as a function of the variables marked by $\dt$ while the remaining points
are fixed parameters.
The expression
$$\mut(\b\Omega,\phi(u^j_1,z+.)\cdots
\phi(u^j_k,z+.)\phi(u^j_{k+1},z)\phi(\ut^{j'}_1,\zt+\dt)\cdots\phi(\ut^{j'}_k,\zt+\dt)\phi(\ut^{j'}_{k+1},\zt)\Omega\k)$$
is a function of the variables marked by a dot (without tilde)
and serves, in turn, as an argument for the functional $\mu$.
Note that for equation \eq{result} to be applicable in the third
and sixth equality, it is necessary that
$$|\zt - z| > \sup_{\zeta\in\D} |\zeta|\quad\mbox{and}\quad \inf_{\zeta\in\D} |\zeta -\zt + z|
> \sup_{\zeta\in\D} |\zeta|\,.$$
This is ensured if the radius $r$ of the disk $\D$ is less than
half the distance $|\zt - z|$.

Following the same approach as in the previous section we try to
make \eq{expanded} arbitrarily small by a suitable choice of the sets $B$ and $\Bt$.
Take a sequence of compact sets $K_m\subset\Md$ such that
$$\Md = \bigcup_{m=1}^\infty K_m\,.$$
For each $m$ we define $B_m$
to be the set of functions
\begin{eqnarray*}
f_{\v{\S\zeta}jj'} & = &
\b\Omega,\phi(u^j_1,z+\zeta_1)\cdots
\phi(u^j_k,z+\zeta_k)\phi(u^j_{k+1},z) \\
& &
\times\,\phi(\ut^{j'}_1,\zt+\dt)\cdots\phi(\ut^{j'}_k,\zt+\dt)\phi(\ut^{j'}_{k+1},\zt)\Omega\k\,,
\end{eqnarray*}
with
$\vz = (\zeta_1,\ldots,\zeta_l)$ running through $K_m$
and $j,j'=1,\ldots,nd^k$. $B_m$ is bounded in $\Fd$.
For a sequence of bounded sets $B_m$ one can find $\rho_m > 0$
such that the union
$$\Bt = \bigcup_{m=1}^\infty \rho_mB_m$$
is again bounded.
This is true in any space described by a countable
family of seminorms (see \cite{Koethe}, p.397).
If $\mut$ is taken from $\Btp$,
$$\max_{1\le j\le nd^k}\max_{1\le j'\le nd^k}\sup_{\v{\S\zeta}\in K_n}|\mut(f_{\v{\S\zeta}jj'})| \le \frac{1}{\rho_n}$$
for each $n\in\bN$. The set $B$ of functions
$$h_{\mut jj'}: \Md\to\bC,\,\v{\zeta}\mapsto
\mut(f_{\v{\S\zeta}jj'})\,,\qquad\mbox{$\mut\in\Btp$, $j,j'=1,\ldots,nd^k$}$$
is therefore bounded in $\Fd$. For $\mu\in\Bp$, $\mut\in\Btp$ and
$\delta = \epsilon^{1/2}/(nd^k)$,
we obtain
\begin{eqnarray*}
|\b\Omega,\phi(\vt_s,\zt)\phi(v_s,z)\Omega\k|
& \le & \sum_{j=1}^{nd^k} |b_j|\sum_{j'=1}^{nd^k}|\bt_{j'}|\,|\mu(\mut(f_{\v{\S .}jj'}))| \\
& \le & \sum_{j=1}^{nd^k}\sum_{j'=1}^{nd^k}\,\delta^2 |\mu(B)|
\\
& \le & (nd^k\delta)^2 =\,\epsilon\,.
\end{eqnarray*}
The inequality remains valid for $v_s$ and $\vt_s$ of the form
\eq{convex_hull}. Thus, we arrive at the result that
two-point blocks $C_2(z,\zt)$ are continuous in $\THud$ if
the open disk $\D$ has radius
$$r < \frac{|\zt - z|}{2}\,.$$

\section{Further Questions}
\label{Further_Questions}
Let us briefly address some of the open questions:
\begin{itemize}
\item Given that 2-point blocks are continuous in Huang's topology (for sufficiently small $\D$), does a corresponding statement hold in the case of Gaberdiel's and Goddard's topologies? The task would be to show that
\begin{equation}
\label{twopointblock}
|\b\Omega,\phi(\vt_s,\zt)\phi(v_s,z)\Omega\k|
\end{equation}
goes to zero when both nets $\{\vt_s\}_{s\in S}$ and $\{v_s\}_{s\in S}$
converge to 0 in the topology $\TGG$ on $V$. The latter means that
for all $\epsilon > 0\,$, compact subsets $K\in\Mo$ and multiindices $I = (i_0,i_1,\ldots,i_l)\in \{1,\ldots,d\}^{l+1}$, there is an index $s_0$ such that
\begin{equation}
\label{convcon}
\|\b x_{i_0},\phi(x_{i_1},.)\cdots
\phi(x_{i_l},.)v_s\k\|_K < \epsilon
\end{equation}
for all $s\ge s_0$ (see sec.\ \ref{Gaberdiels_and_Goddards_Topology} and \ref{Comparison_of_the_Topologies}). The difference to Huang's topology is that here we do not see an obvious way in which the convergence property \eq{convcon} could be used to make estimates on \eq{twopointblock}.
\item What is the relation to the (pre-)Hilbert space topology, i.e.\ the one induced by the inner product? Are $\TGG$ and $\THud$ finer, coarser or neither of it?
\item Are $\THud$ and $\TGGd$ dual to each other? Suppose that the two topologies are defined on linear spaces $\WHu$ and $\WGG$ containing $W$ and let us denote the completions by $\hWHu$ and $\hWGG$ respectively. Duality would mean that $\hWHud = \hWGG$ and/or $\hWGGd = \hWHu$. The question is motivated by a certain correspondence in the way the topologies are constructed. $\WGG$ arises from a projective limit of maps
$$\gd(x_I\otimes .) : \WGG\to F^{\D}_l\,,$$
where, for each $l$, $I$ indexes a basis in the space $\Yb\otimes X^{\otimes l}$ (see \eq{onlyWF} in sec.\ \ref{Gaberdiels_and_Goddards_Topology}). Dualizing these maps gives
\begin{equation}
\label{simpleindlimit}
\gdd(x_I\otimes .) : F^{\D{\textstyle *}}_l\to\WGGd\,,
\end{equation}
and the topology on $\WGGd$ is presumably an inductive limit topology associated to these maps. This resembles the sequence
$$\ed(x_J\otimes .) : \Fdd\to\overline{\rm Hom}(W,\bC)\supset\WHu$$
we obtain from the maps $\ed$ of Huang's topology if $J$ indexes a basis of $X^{\otimes k}\otimes Y$ for each $k$ (see \eq{ed} in sec.\ \ref{Huangs_Topology}). One should note, however, that $\THud$ is a {\it strict} inductive limit and likely to be different from a simple inductive limit as in \eq{simpleindlimit}, even if $\gdd$ and $\ed$ turned out to be equivalent for $l=k$. In that case, it should at least be possible to define alternative topologies that represent the exact duals of $\THud$ and $\TGGd$. A detailed analysis of this issue will appear elsewhere.
\end{itemize}

\end{document}